\documentclass[%
 reprint,
 amsmath,amssymb,
 aps,
]{revtex4-2}
\usepackage{comment}
\usepackage{graphicx}% Include figure files
\usepackage{dcolumn}% Align table columns on decimal point
\usepackage{bm}% bold math
\usepackage{comment}
\usepackage{footmisc}
\usepackage{hyperref}% add hypertext capabilities
\usepackage[mathlines]{lineno}% Enable numbering of text and display math
\begin{document}

\preprint{APS/123-QED}

\title{Comparing the performance of direct and parametric drives for piezoelectric MEMS actuators}

\author{Samer Houri}
\email{samer.houri@imec.be}
\author{Veronique Rochus}

\affiliation{%
 imec, Leuven, Belgium
}%

\date{\today}

\begin{abstract}
This work investigates and compares the response of piezoelectrically actuated nonlinear microelectromechanical devices (MEMS) to direct and to degenerate parametric drives. We describe the regime of degenerate parametric amplification in piezoelectric Duffing-type nonlinear MEMS devices using a single mode expansion, we then explore the existence of regions in parameter space where parametric excitation maybe advantageous compared to direct drive, which we label ``parametric advantage''. Analytical, experimental, and numerical verification demonstrates that parametric advantage can not exist if both pump and signal voltages are accounted for in the total voltage budget. This work determines non-dimensional scaling rules that can act as guidelines for selecting an optimal operating regime for degenerate parametric amplification.%, depending on the application envisioned.
\end{abstract}

\maketitle

%\tableofcontents

%\section{\label{sec:level1}Introduction}

The description of parametric amplification in atomic force microscope cantilever by Rugar et al. \cite{rugar1991mechanical} was the first to demonstrate amplification and noise squeezing in a microelectromechanical device (MEMS). Furthermore, the authors established a simplified analytical expression for parametric amplification in a linear MEMS device, which relates the parametric gain to the quality factor and the parametric modulation depth. Thereafter, several demonstrations of parametric amplification and oscillations followed in a large range of micro- and nano- devices including MEMS resonators \cite{mahboob2008bit,mahboob2008piezoelectrically,gonzalez2018study,prakash2012parametric,turner1998five,carr2000parametric}, nanoelectromechanical device (NEMS) resonators \cite{karabalin2010efficient, collin2011nonlinear,eichler2011parametric,yang2025symmetry}, micromirrors \cite{pribovsek2022parametric,kim2013parametrically}, sensors \cite{wakamatsu2005parametrically,thomas2013efficient}, gyroscopes \cite{harish2008experimental}, and energy harvesters \cite{jia2013multi,beigh2020highly} to name just a few.\\
%\indent\indent\ Both parametric amplification and parametric oscillation have been demonstrated in M/NEMS devices, with each of these regimes possessing their respective points of interest. Mainly, the former has a well-defined amplification factor and an equally well defined bandwidth, whereas the latter regime has a larger oscillation amplitude but its bandwidth and amplitude are ill-defined as these tend to be set by the non-idealities in the system. Thus, even though parametric oscillation produces a larger vibration amplitude%, and has certainly received more attention in the M/NEMS scientific literature
%, parametric amplification remains critical for cases where a precisely controlled amplitude of oscillation, or an accurate amplification gain are required, such as in cases relevant to precision actuators and sensors.\\
%\indent\indent\ This leaves interesting questions: given that for precision applications it is preferable to leverage parametric amplification then how does parametric amplification compares to directly driving the system in question? And would such parametric amplification be more advantageous than directly driving the device? If so, under which conditions?\\
\indent\indent\ In the case of M/NEMS devices, parametric pumping can take on an additional layer of complexity, since M/NEMS devices exhibit (typically cubic) nonlinearities, which only become more pronounced as the devices are scaled down \cite{postma2005dynamic,westra2010nonlinear}. These Duffing-type nonlinearities change significantly the typical responses of parametric systems \cite{aghamohammadi2019response,kumar2011nonlinear,rhoads2010impact,zaghari2016dynamic,neumeyer2017effects,dykman1998fluctuational}.\\% For instance, with the presence of a cubic nonlinearity the optimal amplification phase difference between the signal and the pump (with zero detuning) is no longer $\pi/2$ but amplitude dependent \cite{rhoads2010impact}. Furthermore, upon undergoing parametric oscillations, nonlinear M/NEMS devices can exhibit tri-stable states rather than the typical bistable state \cite{mahboob2008bit,dykman1998fluctuational}.\\% Despite this, a non-dimensional scaling study of the impact of nonlinearities in M/NEMS devices on their performance as parametrically-amplified sensors and actuators remains sought after.\\
\indent\indent\ This work focuses on the regime of degenerate (i.e., frequency of pump = 2 x frequency of signal) parametric amplification and aims to %clarify the issues stated. In particular, we aim to 
answer the following questions; How to express the responsiveness of a nonlinear MEMS device to parametric amplification in terms of the device parameters? Furthermore, given a voltage constrained operation, when does the parametric amplification regime become advantageous compared to directly forcing the system? And how does this ``parametric advantage'' depend on the intrinsic properties, including nonlinearity, of typical M/NEMS resonators?\\
\indent\indent\ The concept of parametric advantage is represented schematically in Fig.~1. A typical resonance mode of a MEMS device, represented by a lumped spring-mass system, is driven via a harmonic force. As this direct forcing is increased the amplitude of the resonator increases, as shown in the figure, for simplicity only the linear case is shown. However, it is equally possible to increase the amplitude of the resonator by periodically tuning the spring's stiffness at twice the drive frequency (i.e., parametric pumping) where the response of the resonator to a small driving force becomes amplified, with the amplification factor increasing asymptotically as the modulation strength approaches the threshold of instability. At a given point the response of the parametrically pumped resonator overtakes that of direct forcing, we label this region ``parametric advantage'' and define it as the region in parameter space where the response of the resonator to parametric amplification %(while accounting for both signal and pump voltage) 
exceeds the response of the resonator to direct forcing at the same overall voltage. If one does not account for the parametric pump in the overall voltage budget, then parametric amplification would be purely advantageous and we obtain the typical parametric gain expressions \cite{rugar1991mechanical,rhoads2010impact}. This traditional approach does not account for the pump voltage in its gain calculations whereas in many examples of sensors and actuators the overall voltage is largely constrained.\\%include the signal in the voltage calculations and it is mostly suited for sensing applications where the signal is externally applied, whereas for precision actuators constrain the overall voltage.\\
%as to whether signal voltage should be included in the total voltage or not. Since this is a matter of definition, we label the mode where voltage constraints applies to both the pump and the signal as the ``actuator mode'' since this would be the most relevant case, and the mode where voltage constraints are solely applicable to the parametric pump as the ``sensor mode'' since in this latter case an external signal is applied.\\
%\indent\indent\ The regime under consideration in this work is that of parametric amplification, therefore the region of instability in parameter space, which corresponds to parametric oscillation, is not the main focus. The motivation behind this choice is due, in addition to the practical application of parametric amplification, to the fact that once parametric amplification sets in, the large amplitude of oscillation in the unstable regime is set by saturation mechanisms that can be attributed to a combination of dispersive and dissipative nonlinearities, which are factors that could be extrinsic to the device itself. For instance, at high amplitudes dispersive nonlinearity can be attributed to read-out saturation [Menno], whereas dissipative nonlinearity can be attributed to pump depletion.\\
\begin{figure}[hbt!]%[p][th]
	\graphicspath{{Figures/}}
	\includegraphics[width=85mm]{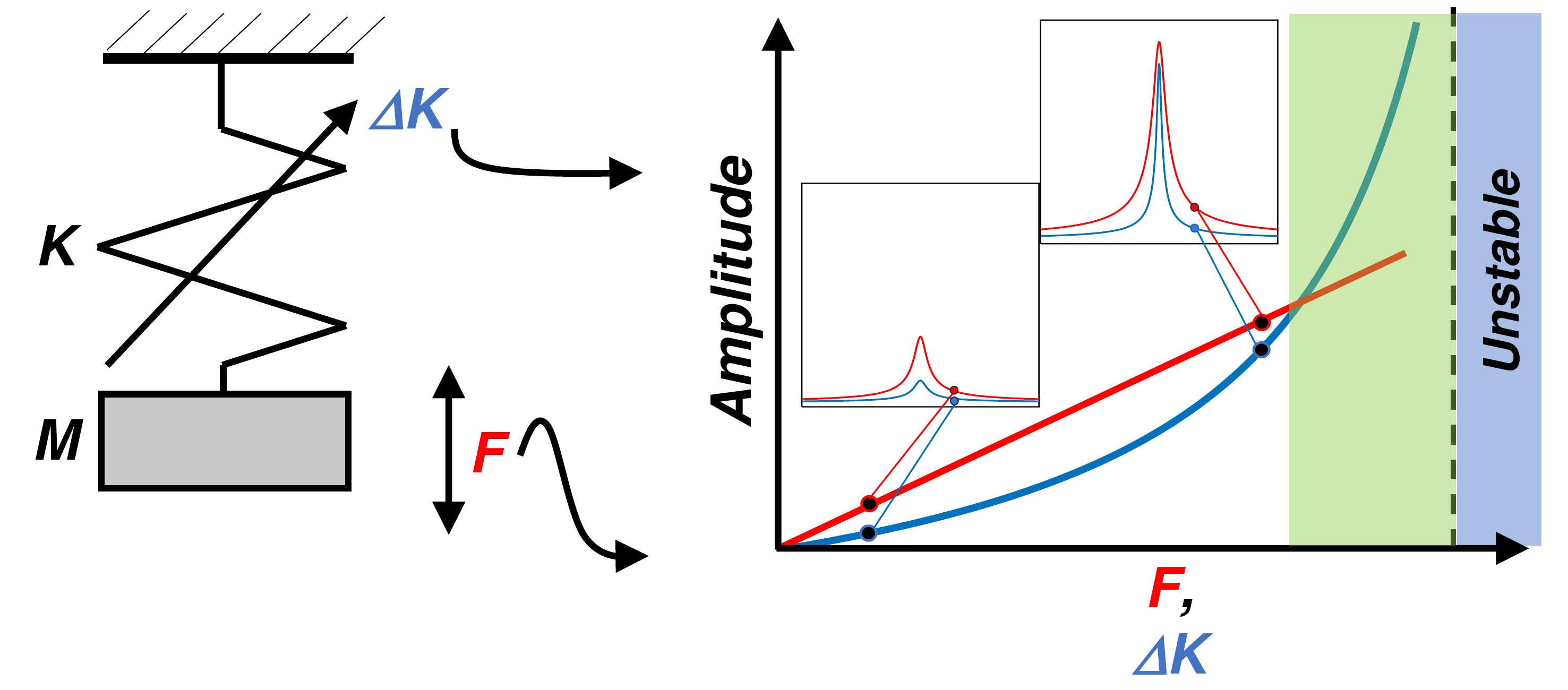}
	\caption{Schematic representation of parametric amplification. If a linear resonator is driven via a harmonic force (\textit{F}) its amplitude increases linearly with forcing (red trace). If it is parametrically pumped by modulating one of its parameters, usually the stiffness ($\Delta K$), the amplitude increases asymptotically (blue trace). The areas of ``parametric advantage'' and parametric instability are shaded in green and blue, respectively. An overlay of the two plots, shown in the insets, demonstrates the narrowing of the line-widths as a function of increased parametric pumping. Note that since the x-axis is an overlay of both \textit{F} and $\Delta K$ care should be taken in expressing them in the correct units.}
	%\caption{Schematic representation of parametric amplification. A linear resonator can be driven via a harmonic force (\textit{F}) in which case its amplitude increases linearly with forcing (red trace), or it can be parametrically pumped by modulating one of its parameters, usually the stiffness ($\Delta K$), in which case the amplitude increases asymptotically (blue trace). An overlay of the two plots, shown in the insets, the area of parametric advantage (green) where the amplitude of a parametrically pumped resonator exceeds that of a forced resonator. Note that since the amplitude depends simultaneously on both \textit{F} and $\Delta K$ care should be taken in overlaying the two amplitudes on the same axis.}
\label{fig1}
\end{figure}
\indent\indent\ To analytically identify the response of a MEMS device to parametric pumping as a function of device parameters and identify whether regions of parametric advantage exists, a single mode equation of motion is used \cite{rugar1991mechanical}, note that if higher-order modes overlap the pump frequency %(which is not the case here) 
then a multi-mode treatment of the system becomes necessary \cite{houri2020demonstration}. Thus, the single degree-of-freedom equation of motion is expressed as\\
\begin{equation}
    {\ddot{x} + {\gamma}\dot{x} + \omega_0^2 x +  \beta x^2 +  \alpha x^3 = F(t)}
    \label{eqn:Eq1}
\end{equation}
where  $x$ is the modal displacement, and $\gamma$, $\omega_0$, $\beta$, $\alpha$ are respectively the modal linear damping, natural frequency, quadratic and cubic (Duffing) nonlinearities, and $F(t)$ is the driving force combining direct and parametric terms. Note that the resonator is considered to be weakly damped and weakly nonlinear, a reasonable assumption for MEMS devices, therefore $(\gamma/\omega_0,~\beta/\omega^2_0,~\alpha/\omega^2_0) << 1$, and that the force term is $F(t)= \chi V(t)$ with $V(t)$ being the applied voltage and $\chi$ a linear transduction parameter in units of ($ms^{-2}V^{-1}$).\\
\indent\indent\ Since the driving force contains two frequency terms, the direct drive and the parametric pumping, it can be expressed as $F(t) = F_D\cos(\omega t) + F_P\cos(2\omega t + \theta)$, where $F_D$ is the direct drive term, $F_P$ is the parametric pumping term, $\omega$ is the direct drive frequency, $2\omega$ is the parametric pump frequency, and $\theta$ is any eventual phase difference between the direct and parametric voltages. It is then equally desirable to expand the displacement ($x$) into two frequency components such that $x = (A_1e^{iwt} + A_1^*e^{-iwt})/2 + (A_2e^{i2wt} + A_2^*e^{-i2wt})/2$, where the star symbol denotes complex conjugate, and $A_1$ and $A_2$ are the complex amplitudes of the $1\omega$ and $2\omega$ components, respectively. Furthermore, since the separation between the two frequency components is large (i.e. $=\omega$), it can be safely presumed that no chaotic response will set in \cite{houri2020generic} for reasonable forcing amplitudes.\\
\indent\indent\ Placing the expanded form of $x$ in Eq.~(1), and introducing a detuning parameter ($\delta$) such that $\omega = \omega_{0}\times(1+\delta)$, gives a coupled complex-amplitude algebraic equation (see supplementary material for detailed derivation) which reads\\
\begin{eqnarray}
\begin{cases}
{\left(\frac{3\bar{\alpha}}{8}(2\left|A_2\right|^2 + \left|A_1\right|^2) -\delta + i\frac{\bar{\gamma}}{2}\right)A_1 + \frac{\bar{\beta}}{2}A_2A_1^* = \frac{\bar{F}_D}{2}}\\
{A_2  = -\frac{\bar{F}_P}{3}e^{i\theta}}
\label{eqn:Eq2}
\end{cases}
\end{eqnarray}
\indent\indent\ Note that in Eq.~(\ref{eqn:Eq2}) a steady-state solution is sought, i.e., $\dot{A_1}=\dot{A_2}=0$, and that nonlinear terms in the $A_2$ equation are negligible. Furthermore, in Eq.~(2) the parameters are expressed in a normalized form with $\bar{F}_D \rightarrow F_D/\omega_0^2$, $\bar{F}_P \rightarrow F_P/\omega_0^2$, $\bar{\chi} \rightarrow {\chi}/\omega_0^2$, $\bar{\gamma} \rightarrow \gamma/\omega_0$, $\bar{\alpha} \rightarrow \alpha/\omega_0^2$, and $\bar{\beta} \rightarrow \beta/\omega_0^2$.\\
\indent\indent\ Equation~(\ref{eqn:Eq2}) demonstrates that the quadratic nonlinear term $\bar{\beta} A_2$ plays the role of parametric pump terms in parametrically excited MEMS/NEMS resonators \cite{rugar1991mechanical}. Therefore, no single-mode (and single degree-of-freedom) parametric pumping can take place without the presence of a quadratic nonlinear term. Furthermore, expressing the equation as a function of $A_2$ allows to relate the parametric pumping term (i.e., $\bar{\beta} A_2$) to the device mechanical or electromechanical properties \cite{emam2004nonlinear, younis2011mems}. If we write $A_1 = \left|A_1\right|e^{\phi}$ and introduce $E = \left|A_1\right|^2$, we can rewrite equation~(\ref{eqn:Eq2}) in a closed form that is independent of $A_2$ which reads\\
\begin{equation}
    {\left(\frac{3\bar{\alpha}}{8}E -\delta'\right)^2E + (\frac{\bar{\gamma}'}{2})^2E  = \frac{\bar{F}_D^2}{4}}\\
    \label{eqn:Eq3}
\end{equation}
\indent\indent\ where $\delta '=-\frac{\bar{\alpha} \bar{F}_P^2}{4} +\delta +\frac{\bar{\beta} \bar{F}_P}{6}\cos(\theta-2\phi)$, and $\bar{\gamma'}=\bar{\gamma}-\frac{\bar{\beta} \bar{F}_P}{3}\sin(\theta-2\phi)$. The above equation simply states that for a sub-threshold degenerate parametric amplification of a nonlinear M/NEMS device, the system will continue to act as a Duffing type resonator \cite{cleland2013foundations}, but with a modified detuning and damping.\\
%\indent\indent\ Since parametric advantage is defined as the threshold upon which parametric drive gives a larger response than direct drive for the same voltage, we pursue two interpretation of this definition. The first accounts for the total applied voltage, wherein we are interested in accounting for both the signal and the pump voltages for the case of parametric amplification. This case is most relevant for actuators [mirrors] (and will be referred to from hereon as the actuator case) where the interest is in obtaining the highest possible amplitude for a give total operating voltage, and can be written down in a more rigorous form as $\eta=A_P(F_D(V_S), F_P(V_P))/A_D(F_D(V_T))) \ge 1$, where $\eta$ is the ratio of the parametrically driven to the direct-only drive case, and $A_D$ and $A_P$, are the amplitudes for the direct and parametric drive cases, and $F_D$, $F_P$ are the direct and parametric forcing terms, and $V_T$, $V_S$, and $V_P$ are the total applied voltage, the signal voltage and the parametric pump voltage, respectively. Note that since we're interested in comparing direct and parametric drives for the same total voltage we have $V_T = V_S + V_P$.\\
\indent\indent\ Now let $V_T$, $V_S$, and $V_P$ be the total permissible voltage, the signal voltage and the parametric pump voltage, respectively. For a voltage constrained actuator case the sum of the pump and signal voltages is limited, thus $V_T = V_S + V_P$. The condition for parametric advantage can be written down in a more rigorous form as $\eta=A_P(V_S,V_P)/A_D(V_T) \ge 1$, where $A_D$ and $A_P$ are the obtained amplitudes for the direct-only and parametric drive cases, respectively, and $\eta$ is the ratio of the two. Note that $A_D$ and $A_P$ should not be confused with $A_1$ and $A_2$, as the former pair are meant to indicate the amplitude (always around $1\omega$) for a system under either direct drive or parametric amplification. Whereas the latter two are meant to indicate the amplitude of the system at $1\omega$ and $2\omega$ respectively regardless of which excitation technique is employed.\\
%\indent\indent\ To obtain the onset of parametric advantage Eq~(2) needs to be solved, this task is however greatly simplified if only a linear system (i.e., $\alpha=0$) is being considered, in such a case $\eta$ takes a closed form of $\eta = A/B$ with $A= F_D/(F_D+F_P)$ and $B=(1-\beta F_P/3\gamma)$. If the Duffing (or higher order dispersive) nonlinearity is present (i.e., $\alpha\neq0$), as is the case here, the expression for $\eta$ becomes too cumbersome to be given in a closed form, but can be solved numerically. However, as will be shown below, the linear case still represents an upper bound on the cases where dispersive nonlinearities are present. It is worth noting that $\eta$ is only dependent on the non-dimensional parameters $A$ and $B$, making the relation given above universal.\\
\indent\indent\ Regions of parametric advantage correspond to the parameter space where $\eta \ge 1$ in Eq.(\ref{eqn:Eq3}). While this will be explored below, we first look at the linear case as it provides valuable insight. Without loss of generality if we consider $(\theta-2\phi)=\pi/2$, for the case of on-resonance driving ($\delta=0$) it is very easy to show that parametric advantage sets in when $V_T = 3\bar{\gamma}/(\bar{\chi}\bar{\beta})$. This $V_T$ value is equivalent to the critical voltage ($V_{cr}$) for the onset of instability (see supplementary material). Counterintuitively, a parametric amplification advantage does not exist for linear M/NEMS when budgeting for both the signal and the pump, not below a total voltage equivalent to the threshold of instability in any case. This conclusion carries when $\delta\neq0$ (with a different $V_T$ value), and the introduction of a Duffing nonlinearity does not change this conclusion.\\%, as will be shown.\\
%On the other hand, the ``sensor case'' which does not accounts for the signal voltage would allow for parametric advantage, although the closed form expression of the condition is complicated.\\
%Eq~(2) needs to be solved, this task is however greatly simplified if only a linear system (i.e., $\alpha=0$) is being considered, in such a case $\eta$ takes a closed form of $\eta = A/B$ with $A= F_D/(F_D+F_P)$ and $B=(1-\beta F_P/3\gamma)$. If the Duffing (or higher order dispersive) nonlinearity is present (i.e., $\alpha\neq0$), as is the case here, the expression for $\eta$ becomes too cumbersome to be given in a closed form, but can be solved numerically. However, as will be shown below, the linear case still represents an upper bound on the cases where dispersive nonlinearities are present. It is worth noting that $\eta$ is only dependent on the non-dimensional parameters $A$ and $B$, making the relation given above universal.\\
%\indent\indent\ Equation~(2) indicates that the $2\omega$ amplitude component ($A_2$) plays the role of a parametric pump thanks to the quadratic nonlinearity. Hence, by plotting the parametric gain as a function of $A_2$ give the same asymptotically increasing relation as shown in a measured example in Fig.~3(a).\\
\indent\indent\ Experimentally, the results are validated using a set of flexural PZT piezoelectric disk MEMS resonators, shown in the inset of Fig.~2(a). The resonators have radii of $150~\mu m$, $200~\mu m$, and $250~\mu m$, with two devices of each diameter measured, and for the case of the $250~\mu m$ devices the second axisymmetric bending mode was equally measured. The devices consist of a stack of 6 micrometer thick layer of structural Silicon on top of which a metal-sandwiched 1 micron thick PZT layer is deposited. Before undertaking measurements, the PZT is poled and electrically stressed using 20 Vdc for a duration of an hour.\\
\indent\indent\ The vibration of the membranes are measured using a Polytec MSA500 laser Doppler vibrometer (LDV) in ambient conditions. An arbitrary waveform generator (KEYSIGHT 33500B) is used to generate the excitation signals which are fed into an $\times 50$ amplifier (Falco WMA-300) before being applied to the resonators via the contact pads using probes. During measurements 4 different Vdc values are applied (10, 15, 20 and 25 V) with the magnitude of the applied ac voltages such that the total voltage does not drop below zero to avoid ferroelectric depolarization \cite{fragkiadakis2022heat}, i.e., Vdc - Vac $> 0$.\\
\indent\indent\ When measured, all devices show easily detectable resonance peaks, Fig.~2(a). The frequency sweeps are fitted to extract the main characteristics of the resonators, i.e., the resonance frequencies, quality factors, and Duffing nonlinearities, which are plotted in Fig.~2(a) through 2(c) as a function of Vdc. Whereas resonance frequencies demonstrate a near linear dependence on Vdc, the quality factor and Duffing nonlinearity do not.% In Fig.~2(c), a large spread is visible in the value of the Duffing parameter for the $250~\mu m$ device for both the first and second modes, this is because the two measured devices come from physically different dies, whereas the other devices are all located next to each other on the same die.
 The drums with the smallest radius (i.e., $R=150~\mu m$) show a negligible Duffing parameter.\\%Unlike the quality factor which also does not exhibit a clear dependence on diameter, the Duffing nonlinearity does, as would be expected from the scaling of M/NEMS devices [Feng roukes].\\
\begin{figure}[hbt!]%[p][th]
	\graphicspath{{Figures/}}
	\includegraphics[width=85mm]{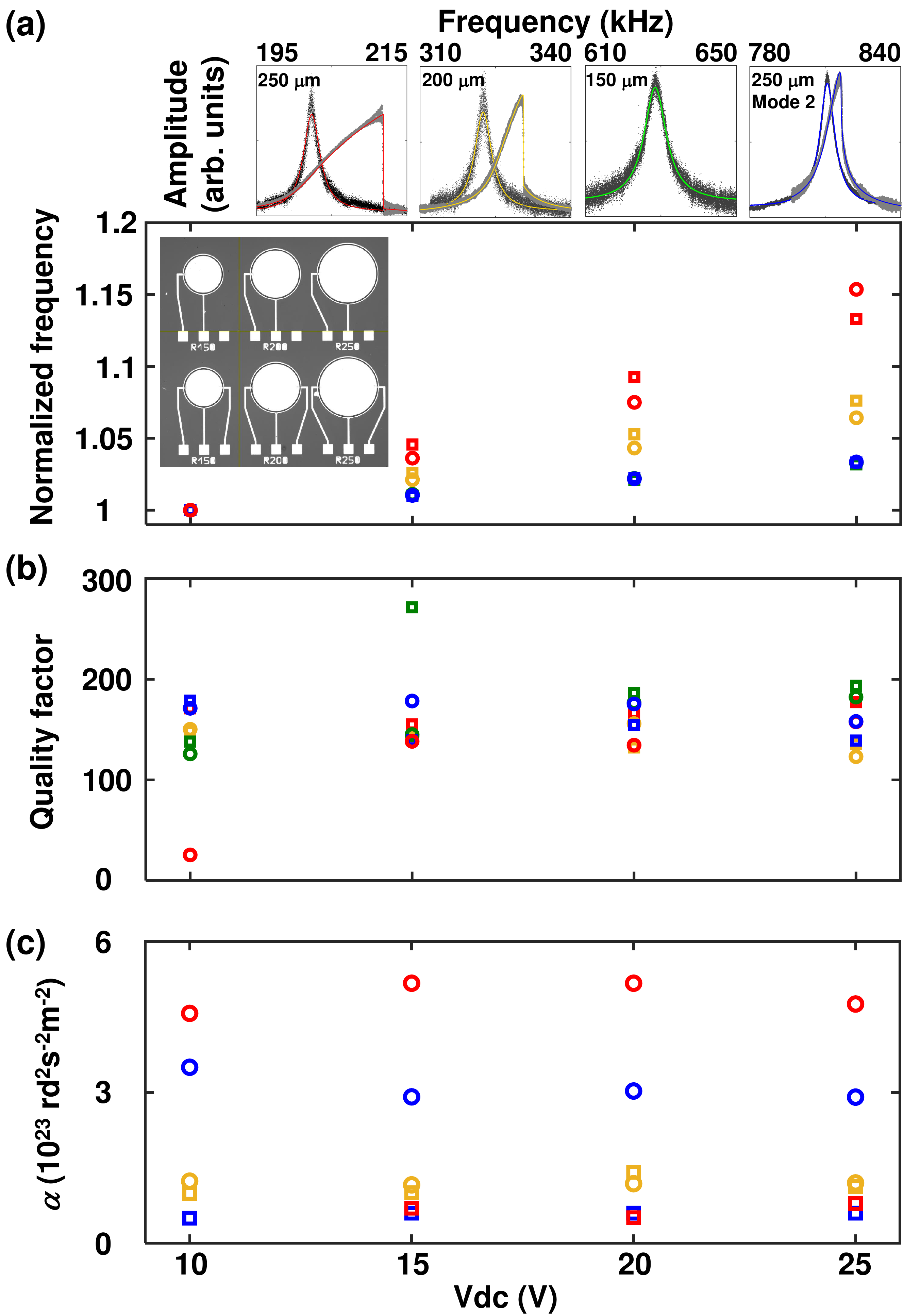}
	\caption{Devices used and their characteristics. (a) Disks of suspended Silicon covered with a metal-sandwiched PZT and having 3 different radii (150, 200, and 250 $\mu m$) are used (shown in inset). (top panels) The linear and nonlinear resonance peaks of these devices (black dots) and their fits (colored lines) are shown as a function of decreasing diameters, with the second mode of a $R = 250~\mu m$ device shown on the rightmost plot (plots obtained for $V_{dc} = 10$~V). The main plot shows the dependence of the modal resonance frequency on dc voltage (two devices for each diameter are used represented by the circle and the square symbols) and demonstrates a near linear dependence of the resonance frequency on $V_{dc}$ for $250~\mu m$ (red), $200~\mu m$ (yellow), $150~\mu m$ (green), and the second mode of the $250~\mu m$ (blue), the frequency values are normalized with respect to the $V_{dc} = 10$~V in order for them to be visible in a single plot. (b) Quality factors as a function of dc voltage with no clear dependence on either voltage or diameter. (c) Duffing parameters as a function of $V_{dc}$ equally showing no clear dependence.}
\label{fig2}
\end{figure}
\indent\indent\ By applying a weak signal to the devices at their respective resonance frequencies along with a strong $2\omega$ voltage, parametric amplification takes place (the phase difference between the two is maintained as $\theta-2\phi=\pi/2$). The amplitudes of vibration at $1\omega$ and $2\omega$ correspond to $\left|A_1\right|$ and $\left|A_2\right|$, respectively. These amplitudes are used to fit Eq.~\ref{eqn:Eq3} and extract the value of the quadratic nonlinearity, i.e. $\beta$, as shown in the example traces in Fig.~\ref{fig3}(a). % Note that the amplitudes are sufficiently small so that the Duffing parameter maybe dropped for the purposes of these fits. 
 The quadratic nonlinear parameters extracted for all the devices and bias voltages are shown in Fig.~\ref{fig3}(b), this parameter equally does not show a strong dependence on the bias voltage.\\
\indent\indent\ The 150 $\mu m$ devices are selected to determine experimentally the response of a linear system to parametric amplification % and confirm the lack of a parametric advantage, particularly
because they lack a measurable Duffing nonlinearity. The devices are driven on resonance ($\delta=0$) with different signal voltage levels ($V_S$) and a swept parametric pump amplitude ($V_P$). Observing the $A_1$ amplitude as a function of the pump voltage gives the typical parametric amplification profile, as shown in Fig.~\ref{fig3}(c) for various drive amplitudes. The actuation and pumping voltages are kept low to avoid damaging the devices, typically below the parametric oscillation threshold (although in some measurements on various diameter devices parametric oscillation threshold was attained). An extrapolation of the direct and parametric drives demonstrates that, as predicted theoretically, the intersection of the two takes place practically upon the onset of instability, thus leaving no room for parametric advantage.\\
\indent\indent\ The same conclusion can be inferred by plotting $\eta$ as a function of $V_S/V_T$ for all dc voltages, as shown in Fig.~\ref{fig3}(d). Knowing that for $\alpha=\delta=0$ we have $\eta = (V_S/V_T)(\bar{\gamma}/\bar{\gamma'})$ (see supplementary material), meaning that the experimental data should fall on a $y=x$ diagonal for no parametric amplification, below the diagonal for the case of de-amplification and above the diagonal for amplification. Visibly, the bulk of the points fall near the diagonal, indicating a relatively weak parametric amplification. More importantly, the data points overwhelmingly fall below the $\eta=1$ limit indicating no parametric advantage. Some points of the data set do cross the limit, but we attribute this to experimental error. Equally, points in the de-amplified regime are attributed to experimental error or limits on resolution, especially in the regions where $V_T/V_S\approx 1$ and $V_S/V_T \approx 0$ (there is a 50 mV shift between the parametric and direct data which becomes relevant at those extremes).\\
\begin{figure}[hbt!]%[p][th]
	\graphicspath{{Figures/}}
	\includegraphics[width=85mm]{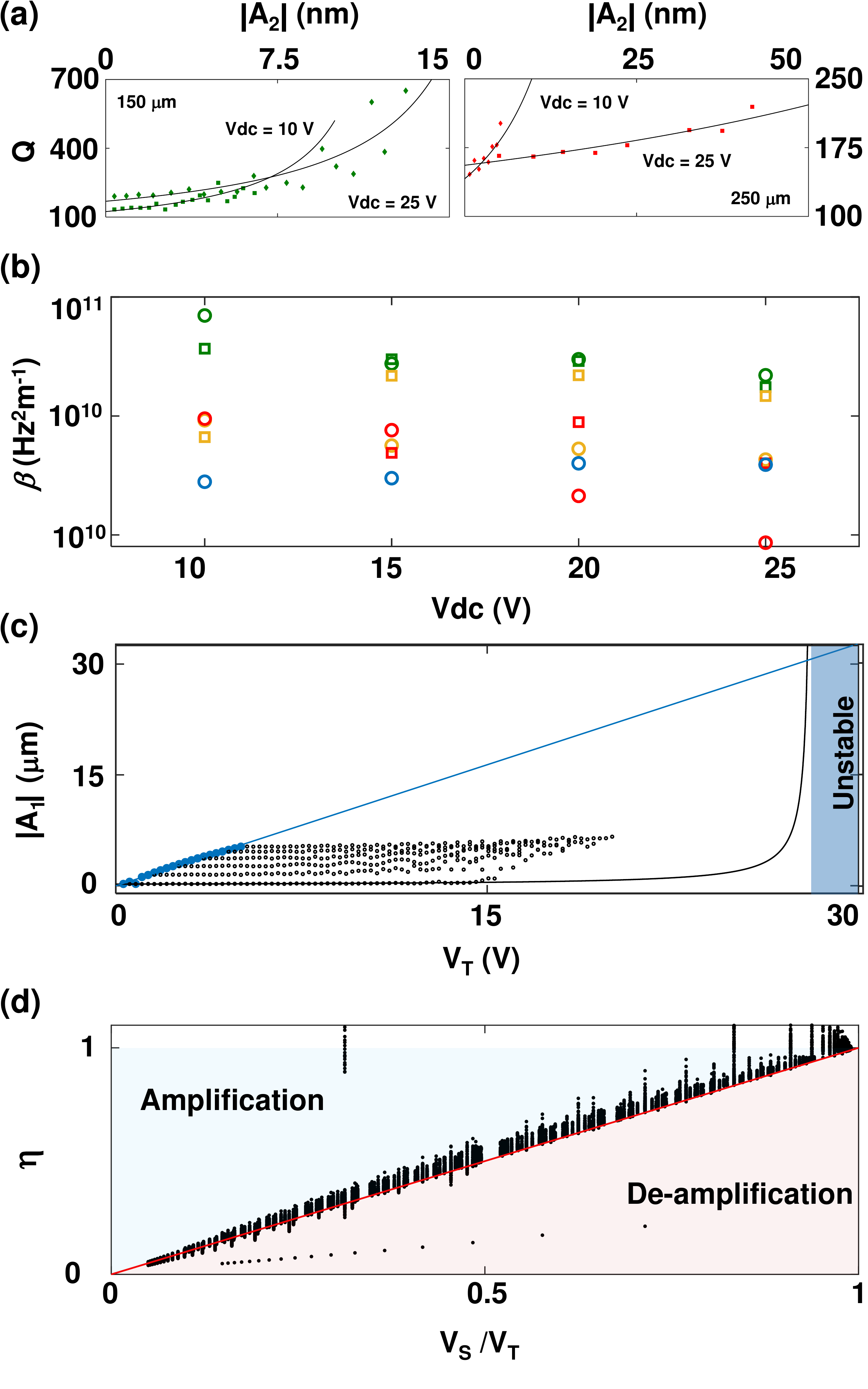}
	\caption{Parametric amplification and quadratic nonlinearity. (a) Example fits of the quality factor versus the $A_2$ amplitude for the 150 $\mu m$ and 250 $\mu m$ devices for $V_{dc} = 10$ and $25$ V. (b) Extracted $\beta$ for all devices showing no strong dependence on dc voltage. (c) Comparison of direct (blue dots) and parametric (black dotes) drives for the 150 $\mu m$ device, for $V_{dc}=10$ V. The plot demonstrates that upon extrapolation the direct drive (blue line) and parametric drive (black line) intersect practically at the onset of instability (blue shaded area). The x-axis is $V_T = V_D$ for the direct drive, and $V_T = V_S + V_P$ for the parametric case. (d) Plot showing the 150 $\mu m$ experimentally measured ratio $\eta$ as a function of the ratio $VS/VT$ (black points), the red line indicates the $y=x$ diagonal, the red shaded area indicates de-amplification region, and the blue shaded area indicates amplification. The data points overwhelmingly fall bellow the $\eta=1$ limit.}  
    \label{fig3}
\end{figure}
\indent\indent\ Determining parametric advantage for the full nonlinear system is not analytically tractable, however, we demonstrate that parametric advantage remains unachievable by solving numerically Eq.~\ref{eqn:Eq3} using the experimentally fitted parameters. To show why parametric advantage does not arise in the case of nonlinear MEMS devices despite the Duffing caused detuning, Fig.~\ref{fig4}(a) shows simulated traces for the directly and parametrically driven 250 $\mu$m device and their ratio $\eta$ obtained for $V_{dc} = 10$ V and an $V_T=1$ and $300$ mV. Fig.~\ref{fig4}(a) shows that at low voltages, parametric pumping plays a negligible role and therefore the signal budget is better spent on directly driving the system. The value of $\eta$ simply reduces to $V_S/V_T$ for the linear case, while a peak in the case of strong drive is visible. This peak can be attributed to Duffing-caused detuning of the resonance peak (the direct and parametric cases are unequally detuned according to Eq.~\ref{eqn:Eq3}). Even with this peak the value of $\eta$ does not surpasses 1. A more detailed calculation is shown in Fig.~\ref{fig4}(b) which shows a 2-dimensional plot of the maximum of $\eta$ calculated numerically for the 250 $\mu$m device for $V_{dc}=10$ V as a function of $V_D$ and $V_P$ normalized to the threshold voltage of instability $V_{cr}=3.24$ V.  Figure~\ref{fig4}(b) demonstrates no possibility for parametric advantage even when a MEMS is driven into the nonlinear regime.\\% Even though these results are obtained numerically, and as such does not constitute a rigorous proof that parametric advantage can not exist in a Duffing-type M/NEMS resonator, however it does exclude the possibility for reasonable values of Duffing and quadratic nonlinearities.\\
\begin{figure}[hbt!]%[p][th]
	\graphicspath{{Figures/}}
	\includegraphics[width=85mm]{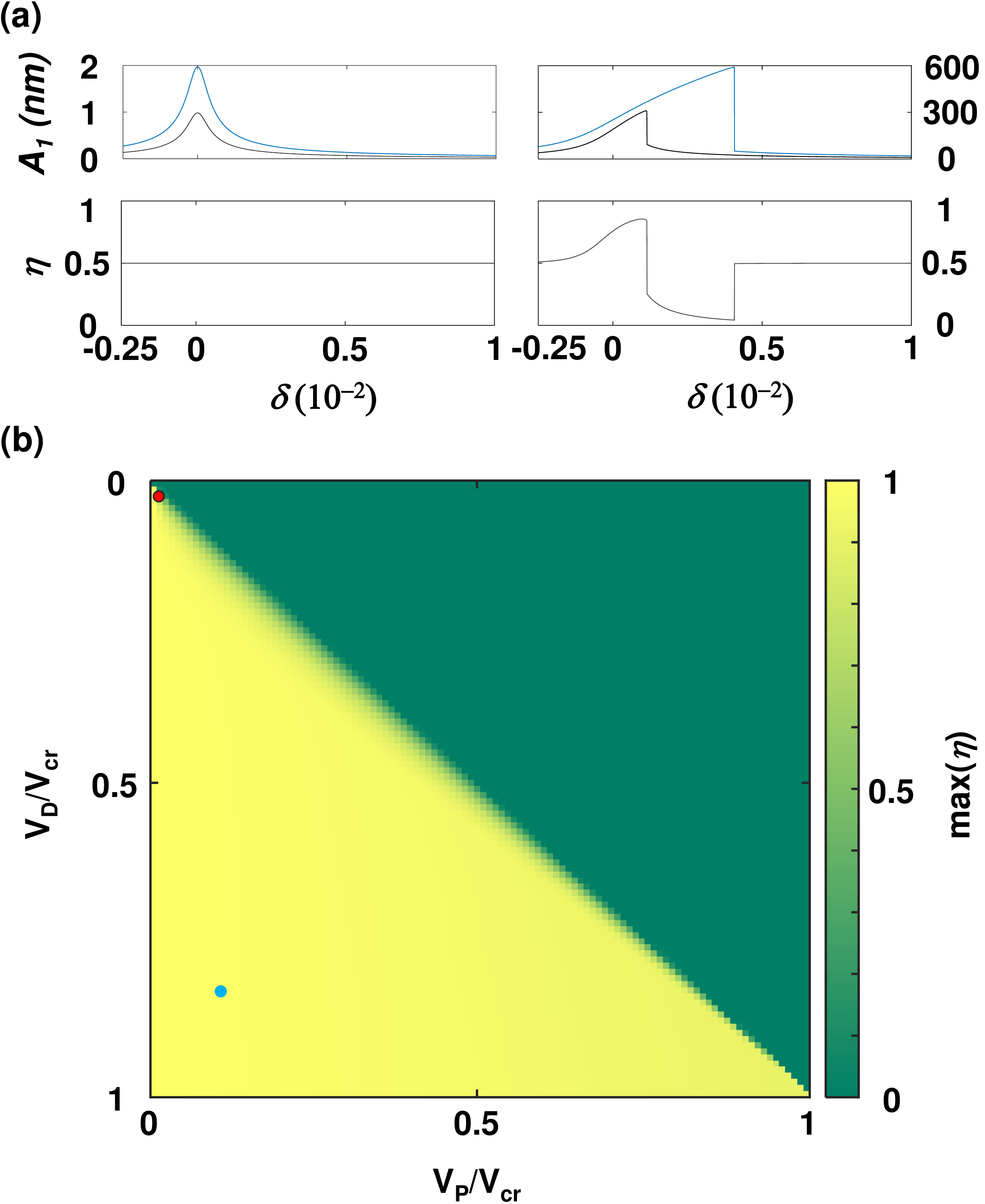}
	\caption{Calculated $\eta$ for the nonlinear case. (a) Two sets of traces for weak linear driving (left) and strong nonlinear driving (right) calculated using the device parameters of the 250 $\mu$m resonator. The linear case, calculated for $V_{dc}=10$ V, $V_{D}=1$ mV, $V_P=0.5$ mV, and $V_S=0.5$ mV, shows $\eta=0.5$. For the nonlinear case, calculated for $V_{dc}=10$ V, $V_D=0.3$ V, $V_P=0.15$, and $V_S=0.15$ V, we see that due to detuning a peak in the value of $\eta$ is obtained, although it does not cross the threshold of 1. (b) The maximum of $\eta$ obtained for calculating all combinations of $V_P + V_S = V_T < V_{cr}$ for the device parameters corresponding to the 250 $\mu$m devices, showing no possibility of parametric advantage. The position in parameter space of the linear and nonlinear traces in (a)  are shown as the red and blue dots, respectively.}  
    \label{fig4}
\end{figure}
%\indent\indent\ Determining parametric advantage for the full nonlinear case is not analytically tractable, however, the task is facilitated by rewriting Eq.~\ref{eqn:Eq3} in the form $E^3 - AE^2 + BE - C =0$, where $A=16\delta'/3\alpha$, $B=64(\delta'^2+0.25\gamma'^2)/9\alpha^2$, and $C=16F_D^2/9\alpha^2$. The $A$, $B$ and $C$ parameters are non-dimensional and universally applicable to any M/NEMS device regardless of the selected detunings, phase difference, or device parameters. The measurements from the various devices and bias voltages can be collapsed onto a 3D plot whose axes are the parameters $A$, $B$, and $C$. By mapping $\eta$ (using a color coding scheme) onto the 3-dimensional figure, it is possible to identify the zone of parametric advantage, as shown in Fig. 4(a) for the ``actuator case'' and~4(b) for the ``sensor case''.\\
\indent\indent\ In summary, this work compared the response of nonlinear M/NEMS devices to direct forcing and parametric amplification using a single-mode expansion. The work described theoretically and demonstrated experimentally the need for a quadratic nonlinear component in order for parametric amplification to take place. Furthermore, this work investigated the regions in parameter space where parametric amplification would be more efficient than direct drive. Surprisingly, such parametric advantage does not exist for linear nor nonlinear, Duffing-type, systems if one accounts for both the pump and the signal%, although it does exist if the signal to be amplified is not considered part of the voltage budget. When we account for the Duffing nonlinearity, we find that regions of parametric advantage do exist even when we account for both the signal and the pump voltages. The interpretation of this fact is the effective detuning that nonlinear M/NEMS devices are subject to when strongly driven. As such, the regions of parametric advantage exist only when the forcing is enough to cause detuning, although in terms of absolute amplitudes the parametrically amplified response remains at best equivalent to the direct drive case
. If, on the other hand, either the signal or the pump voltages are not part of the overall budget then parametric advantage can be achieved.\\
\indent\indent\ See the supplementary material for a derivation of Eq.~\ref{eqn:Eq2}, threshold voltage ($V_T$), and $\eta$.\\
%\indent\indent\ Despite the fact that parametric advantage as defined in this work is not accessible for M/NEMS devices, this certainly does not preclude the usefulness of parametric amplification in driving actuators, for instance, we have limited the analysis to a total voltage below the threshold voltage, i.e., $V_T<V_{cr}$, this condition depends purely on the use case and a higher voltage may be acceptable if the priority is to maintain a desired frequency response.\\
%\indent\indent\ Although the results shown here are related to piezoelectrically actuated MEMS resonators, the conclusions maybe extended to other types of actuation. Indeed, as long as the underlying assumptions are valid, the conclusions maybe even extended to different types of parametric amplifiers such as the Josephson Parametric Amplifier. The results outlined and demonstrated in this work will act as a useful guideline for operating M/NEMS sensors and actuators.%, as it enables to quickly identify which of the direct or parametric drives would be more advantageous, and under which conditions.\\


\providecommand{\noopsort}[1]{}\providecommand{\singleletter}[1]{#1}%

\end{document}